\documentclass[aip,preprint]{revtex4-1}
\usepackage{amsmath,amssymb}

\usepackage{graphicx}

\usepackage{color}

\renewcommand{\vec}{\mathbf}
\newcommand{\bea}{\begin{eqnarray}}
\newcommand{\eea}{\end{eqnarray}}
\newcommand{\vk}{\vec{k}}
\newcommand{\bsm}{\begin{smallmatrix}}
\newcommand{\esm}{\end{smallmatrix}}

\begin{document}

\title{EFFECT OF ION POLARIZATION ON LONGITUDINAL EXCITATIONS \\ IN IONIC MELTS}

\author{B.~Markiv}
\affiliation{Institute for Condensed Matter Physics NAS of
Ukraine,\\ 1~Svientsitskii Str., 79011 Lviv, Ukraine}
\author{M.~Tokarchuk}
\affiliation{Institute for Condensed Matter Physics NAS of
Ukraine,\\ 1~Svientsitskii Str., 79011 Lviv, Ukraine}
\affiliation{Lviv Polytechnic National University,
\\12~Bandera Str., 79013 Lviv, Ukraine}

\date{\today}

\begin{abstract}
A simplified model for a collective dynamics in ionic melts is proposed for
the description of optic-like excitations. Within a polarization model of ionic melt
the analytical expressions for optic and relaxation dipole modes are obtained.
The considered model allows one to describe a softening of frequency and
an increase of damping of optic modes caused by polarization processes in comparison with the rigid-ion model. The contributions related with ion polarization
to time correlation functions are calculated.
\end{abstract}


\maketitle

\section{Introduction}

Despite a long history of investigation of ionic melts \cite{March,Hansen,Price,Par,Gia,Bosse,Tok},
they remain an interesting object of study in different aspects \cite{React,Wang,Perera}.
One of the important tasks is the study of collective dynamics in ionic melts since
they possess their own features in comparison with binary mixtures of neutral particles
\cite{bryk-05,Kuporov}.
Recent experiments on inelastic X-ray scattering in the melts of NaCl, NaI, CsCl \cite{Demmel-04,Demmel-05,Inui-07}
renew  an attention and an interest to the collective dynamics in molten salts.
In particular, in experiments for NaI \cite{Demmel-05} the optic excitations were
observed on the shape of total dynamic structure factor.
Such optic excitations were predicted by means of molecular dynamics (MD) simulations.
However, computer simulations reveal an importance of taking into account the
polarization processes related with an electronic structure of ions.
In the earlier investigations of ionic melts the potentials in the Mayer-Huggins or Tosi-Fumi form
were used. These potentials correspond to the model of rigid ions (RI) and do not take into account
polarization effects. However, as it was shown by means of molecular dynamics  \cite{Jacucci,31,Alcaraz}
and ab initio molecular dynamics (AI MD) \cite{32,bryk,26,klev} simulations, the effects caused by
ion polarization can play an important role.
In particular, the RI models of ionic melt provide different frequency
of optic modes in comparison with AI MD \cite{bryk,26}.
That is why, outer electron shells should be considered with an effective interaction
since they can polarize.

Nowadays there exist several models which allow one to take into account polarization processes in the system.
Within the framework of polarization models \cite{31} the induced point dipoles are added on the ionic sites.
These models were successfully used in the study of the structure and the dielectric properties of
NaI, AgI, AgCl melts \cite{Bit,Bit2,Alcaraz}. Within the shell model \cite{Dixon} used in the study
of NaCl melt, the dipoles are supposed to be of finite length and are modelled by a point
positive charge of a core and a negative one of a shell, which are connected with a harmonic bond.
Fluctuation models \cite{22,Wilson} operate with point fluctuating charges fixed at certain sites.

Recently, based on the polarizable point dipole model the spectrum of collective modes in
ionic melts was theoretically investigated in the limit of small wave vector in Ref.\cite{markiv},
where the contributions to acoustic collective excitations due to
ion polarization were obtained. It was shown that the influence of the
polarization processes over the sound velocity is negligible, however, the sound damping
coefficient is renormalized due to these effects, and this is in agreement with ab initio
molecular dynamics results \cite{23,bryk,26}. The influence of ion polarization on the corresponding
relaxation modes was obtained within the proposed model of collective dynamics as well.

An importance of the polarization effects can be estimated by the shape of
a static concentration structure factor \cite{bryk,31}, which in the long-wavelength
limit has the asymptotics $S_{cc}(k)|_{k\to 0}\sim \frac{\varepsilon k_BT}{4\pi n e^2}k^2$.
If the value of high-frequency dielectric permittivity $\varepsilon$ is about 2 or higher
the polarization effects are essential, whereas for the rigid-ion model $\varepsilon=1$.
In the works \cite{31,klev} it was shown that the polarization processes lead to a sufficient redistribution of the contributions from collective modes to concentration dynamic structure factor.
Herewith, the contribution from the relaxation mode related with the conductivity of the system increases
whereas the one from optic modes decreases in comparison with the rigid-ion model.

Another interesting object of study is optic excitations in ionic melts.
Here the importance of consideration of polarization effects is clearly defined.
In particular, as it was shown by means of computer simulations,
the RI models of ionic melt provide an overestimated frequency
of optic modes in comparison with AI MD \cite{bryk,26} in the region of small wave vectors.
And vice versa, the coefficient of damping of optic modes is underestimated in RI models.
It is available a number of works which report the results of computer simulations \cite{31,26,klev,bryk}, however, there is a lack of studies devoted to the theoretical description of the problem. That is why
in the present paper we are focused on the theoretical study of the influence of ion polarizability on the longitudinal optic excitations in ionic melts. We will propose a simplified model of collective dynamics which allows us to take into account polarization effects and to describe the softening of frequency of optic modes in comparison with the model of rigid ions. The structure of the paper is the following. In the second section we present basic relations of the polarization model of the ionic melt and the approach of generalized collective modes (GCM).
In Sec.~3 we propose the model of collective dynamics, within its framework we obtain the analytical expressions for collective modes and time correlation functions. Here we also provide some quantitative comparison of our results with the data available from computer simulations. In the last section we
summarize the obtained results.

\clearpage

\section{Model and method}

\subsection{Ion-polarization model of ionic melt}

In our study we use the ion-polarization model of the ionic melt which can be presented as follows. The system consists of $N_a$ classical ions of species $a$ ($a=1,2$ or $+,-$) with a charge $Z_a e$ and an induced dipole moment. In general case, both negatively and positively charged ions
can polarize. However, in the case of a number of ionic  melts the polarizability of anions
is much more than the polarizability of positively charged ions due to a larger number of electrons on the outer shell. Due to the interaction with an electric field $\vec E_j$, the dipole moment $\vec d_j^a$ induces on the $j$-th ion:
\bea
\label{1.1}
\vec d_j^a = \alpha_j\vec E_j -
{\alpha_j\sum_b
\sum_{\bsm
i=1 \\ i\neq j
\esm}^{N_b} f_{ab}(r_{ij}) \frac{Z_{b} e}
{r_{ij}^3}\vec r_{ij}}\, ,
\eea
where $\alpha_j$ is a polarizability of the $j$-th ion in electric field
\bea
\label{1.2} && \vec E_j = \vec E_j^d + \vec E_j^q, \\\nonumber
&&\vec E_j^q = \sum_a\sum_{\bsm
i=1 \\ i\neq j
\esm}^{N_a}\frac{Z_{a} e} {r_{ij}^3}\vec
r_{ij}, \\\nonumber
&& \vec E_j^d = \sum_a\sum_{\bsm
i=1 \\ i\neq j
\esm}^{N_a}\left[3\frac{(\vec d_i^a\cdot\vec r_{ij})} {r_{ij}^3}\vec r_{ij} -
\frac{1} {r_{ij}^3}\vec d_{i}^a\right].\\
\eea
Here, $\vec E_j^q$ denotes the field induced by all the charges except the $j$-th one,
and $\vec E_j^d$ is the field due to all dipole moments except the $j$-th one,
$r_{ij}=|\vec r_{i}-\vec r_{j}|$  is the distance between ions.
$f_{ab}(r_{ij})$ is the damping dispersion function \cite{42}.

As a result of the interaction between dipoles and between charges and dipoles,
the dipole moment of ion changes and leads to their rotational motion.
Thus, the Hamiltonian of the system can be presented in the following form:
\bea
\label{1.3}
 H=\sum_a\sum_{j=1}^{N_a}\left(\frac{p_j^2}{2m_a}
 +\frac{1}{2}{\vec
 w}_j^\mathsf{T}{{\mathbf{J}}}_j
{\vec w}_j\right) +U_\mathrm{ion} \,.
\eea
Here, $\vec p_j$ is a momentum of the $j$-th ion, $m_a$ is a mass of an ion of species
$a$, $\vec w_j$ is an angular velocity ($\vec w_j^\mathsf{T}$ stands for transposed vector).
${{\mathbf{J}}}_j$ denotes an inertia tensor of the $j$-th polarized ion (ionic dipole) determined relatively its center of mass.
$U_\mathrm{ion}$ can be presented in the following form \cite{31}:
\bea
\label{1.4}
U_\mathrm{ion}&=&\frac{1}{2}\sum_{a,b}\sum_{\bsm
i,j=1 \\ i\neq j
\esm}^{N_a,\, N_b}\Phi_{ab}(r_{ij})-\sum_a\sum_{j=1}^{N_a}\vec
d_j^a\cdot\vec E_j^q - \frac{1}{2}\sum_a\sum_{j=1}^{N_a}\vec
d_j^a\cdot\vec E_j^d\nonumber
\\ &&+ \sum_a\sum_{j=1}^{N_a}\frac{(d_j^a)^2}{2\alpha_j} +
{\sum_{a,b}\sum_{\bsm
i,j=1 \\ i\neq j
\esm}^{N_a,\, N_b}f_{ab}(r_{ij})
\frac{Z_{b}e}{r_{ij}^3}\vec r_{ij}\cdot\vec d_i^a}
\,.
\eea
$\Phi_{ab}(r_{ij})$ denotes an ion--ion interaction potential:
\bea
\label{1.5} \Phi_{ab}(r_{ij}) = \frac{Z_{a}Z_{b} e^2} {r_{ij}}
+\varphi^{sh.r.}_{ab}(r_{ij}),
\eea
where $\varphi^{sh.r.}_{ab}(r_{ij})$ is a short-range potential containing an overlap repulsive part. Its form depends on a system considered \cite{41,Parrinello}.
$Z_{a}$, $Z_b$ are the valences of the ions of the corresponding species, $e$ denotes an electron charge. The second term in Eq. (\ref{1.4}) is from the ``dipole--charge'' interactions,
the third one is from the ``dipole--dipole'' interactions.
The fourth term denotes the energy necessary to create $N=\sum_a N_a$ induced dipoles \cite{Ahlstrom}.
The last term corresponds to the short-range damping polarization.

\subsection{Generalized collective modes approach}

In order to study the effect of ion polarization on optic excitations we use
the approach of generalized collective modes \cite{40}.
This approach is based on the equations for time correlation functions which can be
obtained by means of various methods, in particular by the Zubarev nonequilibrium statistical operator method \cite{Zub2} or by the Mori projection operator method \cite{mori}.
In the Laplace representation these equations can be written down as follows:
\bea
\label{1.6}
\left\{z\mathbf{I}-i\mathbf{\Omega}(k)
+\tilde{\boldsymbol{\varphi}}(k,z)\right\}\tilde{\mathbf{F}}(k,z)=\mathbf{F}(k,0),
\eea
where
\bea
\label{1.6'}
\tilde{{F}}_{lm}(k,z)=\int_0^{\infty}e^{-zt}{F}_{lm}(k,t)dt\eea
are the Laplace transform of the time correlation function
\bea
\label{1.7}
{F}_{lm}(k,t)=\langle \hat{a}_{l,\vk} \, e^{-iL_Nt} \, \hat{a}_{m,-\vk}, \rangle
\eea
built on the dynamic variables $\{\hat{a}_{l,\vk}\}$. Here, $iL_N$ is the Liouville operator corresponding to the Hamiltonian of the system.
In Eq.~(\ref{1.6}) $\mathbf{I}$ denotes a unit matrix, $i\mathbf{\Omega}(k)$ is a frequency matrix and $\tilde{\boldsymbol{\varphi}}(k,z)$  is a matrix of memory functions.
In the Markovian approximation for the memory functions
it is convenient to write down the equations for
time correlation functions in the following form
\begin{eqnarray}
\label{1.8}
\left\{z\mathbf{I}+\mathbf{T}(k)\right\}\tilde{\mathbf{F}}(k,z)=\mathbf{F}(k,0),
\end{eqnarray}
where
\bea
\label{1.9}
\mathbf{T}(k)=-i\mathbf{\Omega}(k)+\tilde{\boldsymbol{\varphi}}(k,0)
\eea
is the generalized hydrodynamic matrix.
The problem of finding of collective modes in the system reduces to the eigenproblem of  matrix $\mathbf{T}(k)$:
\bea
\label{1.10}
\mathbf{T}(k)\mathbf{X}_\alpha(k)=z_\alpha(k)\mathbf{X}_\alpha(k),
\eea
where $\mathbf{X}_\alpha(k)$ are the eigenvectors corresponding to the eigenvalues $z_\alpha(k)$.
When the eigenvalues and eigenvectors of the matrix  $\mathbf{T}(k)$ are known
the solution of Eq.~(\ref{1.8}) for the Laplace transforms of the time correlation functions (\ref{1.7}) can be written as follows:
\begin{eqnarray}
\label{1.11}
\tilde{{F}}_{lm}(k,z)=\sum_{\alpha}\frac{{G}_\alpha^{lm}(k)}{z+z_\alpha(k)}.
\end{eqnarray}
Here, ${G}_\alpha^{lm}(k)$ are the complex amplitudes defining the contribution from each collective mode to the time correlation functions:
\begin{eqnarray}
\label{1.12}
{G}_\alpha^{lm}(k)=\sum_n {X}_{l,\alpha}(k)[{X}^{-1}(k)]_{\alpha,n}{F}_{nm}(k,0).
\end{eqnarray}
Then, within the framework of the GCM approach the time correlation functions can be presented as a sum of the contribution from corresponding collective excitations
\begin{eqnarray}
\label{1.13}
{F}_{lm}(k,t)=\sum_{\alpha}{G}_\alpha^{lm}(k)e^{-z_\alpha(k)t}.
\end{eqnarray}
This is a great advantage of the method because it allows one to select and to investigate the processes in the system which are of particular interest. We will use this advantage in the study of optic excitations in ionic melts.

\section{Collective modes and time correlations functions}

\subsection{Simplified model of collective dynamics}

In the region of small wave vectors the dispersion of longitudinal propagating collective excitations in the ionic melts can be well described using the dynamic variables of the longitudinal components of total mass current (momentum)  $\hat{J}^t_{\vk}$ and charge current $\hat{J}^q_{\vk}$ (in contrast to the region of large values of wave vector, where the collective dynamics is properly described in terms of partial currents) \cite{bryk}.
Therefore, to study the longitudinal excitations the set of variables should contain these operators as well as theirs first time derivatives:
\bea
\label{1}
{\cal A}^{(4)} = \{ \hat{J}^t_{\vk}, \dot{\hat{J}}^t_{\vk}, \hat{J}^q_{\vk}, \dot{\hat{J}}^q_{\vk} \}.
\eea
Here, the total mass and charge currents, respectively
\bea
\hat{J}^t_{\vk}=\hat{J}_{1,\vk}+\hat{J}_{2,\vk},
\qquad\qquad
\hat{J}^q_{\vk}=\frac{Z_1e}{m_1}\hat{J}_{1,\vk}+\frac{Z_2e}{m_2}\hat{J}_{2,\vk},
\eea
are expressed via the partial currents:
\bea
\hat{J}_{a,\vk}=\sum_{l=1}^{N_a} p_l^a e^{i\vk\cdot\vec{r}_{l}}
, \qquad a=1,2 \quad \text{or} \quad +,- \ .
\eea
In order to take into account the polarization processes we introduce the corresponding dynamic variable
\bea
\label{4}
\hat{{d}}_{\vk}^{a}=\sum_{l=1}^{N_a}
{d}_{l}^{a}e^{i\vk\cdot\vec{r}_{l}},
\eea
which describe the induced dipole moment ${d}_{l}^{a}$ Eq. (\ref{1.1})
of the ions of species $a$.

\begin{figure}[!t]
\centerline{
\includegraphics[width=0.5\textwidth]{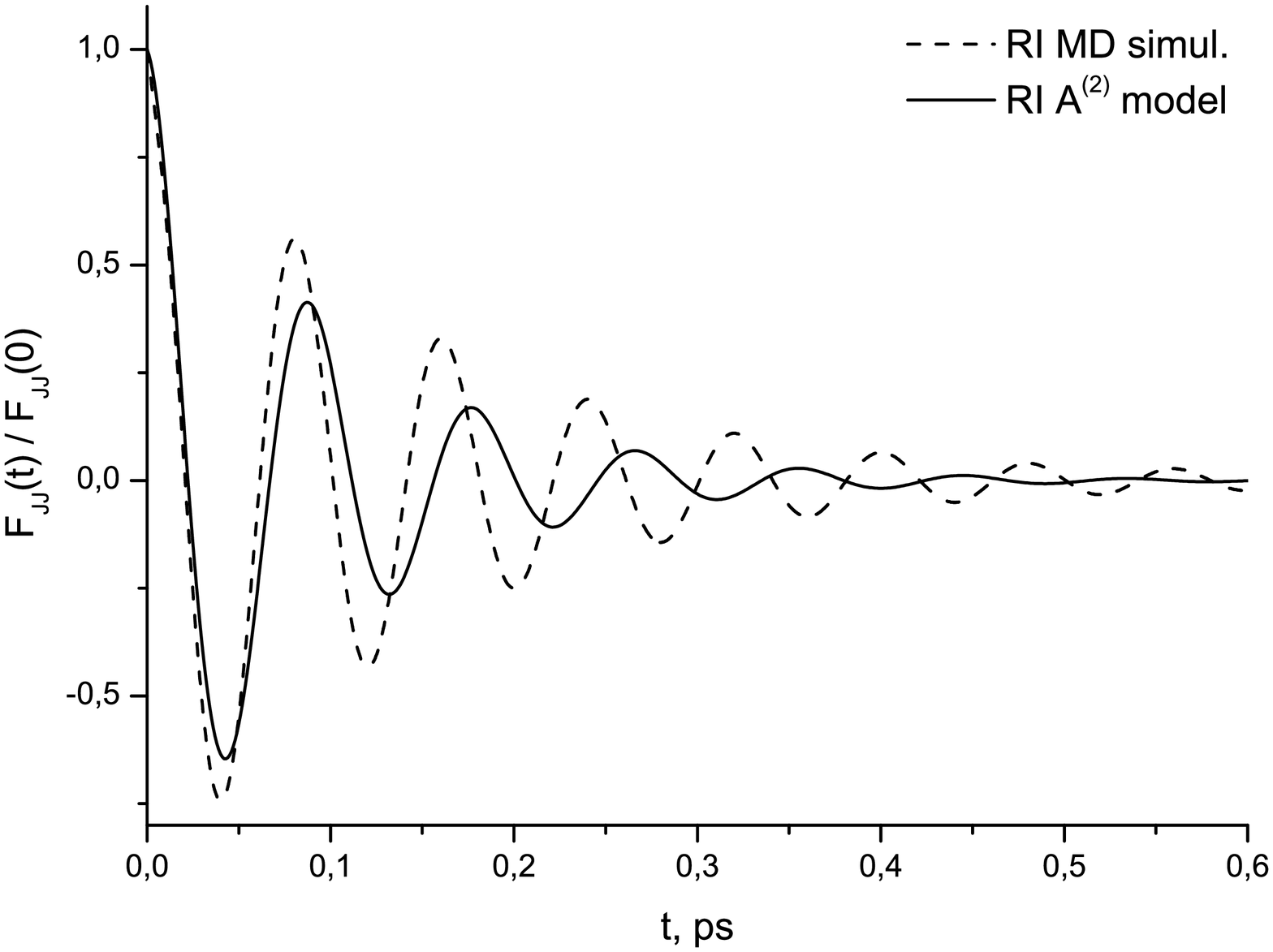}
\includegraphics[width=0.5\textwidth]{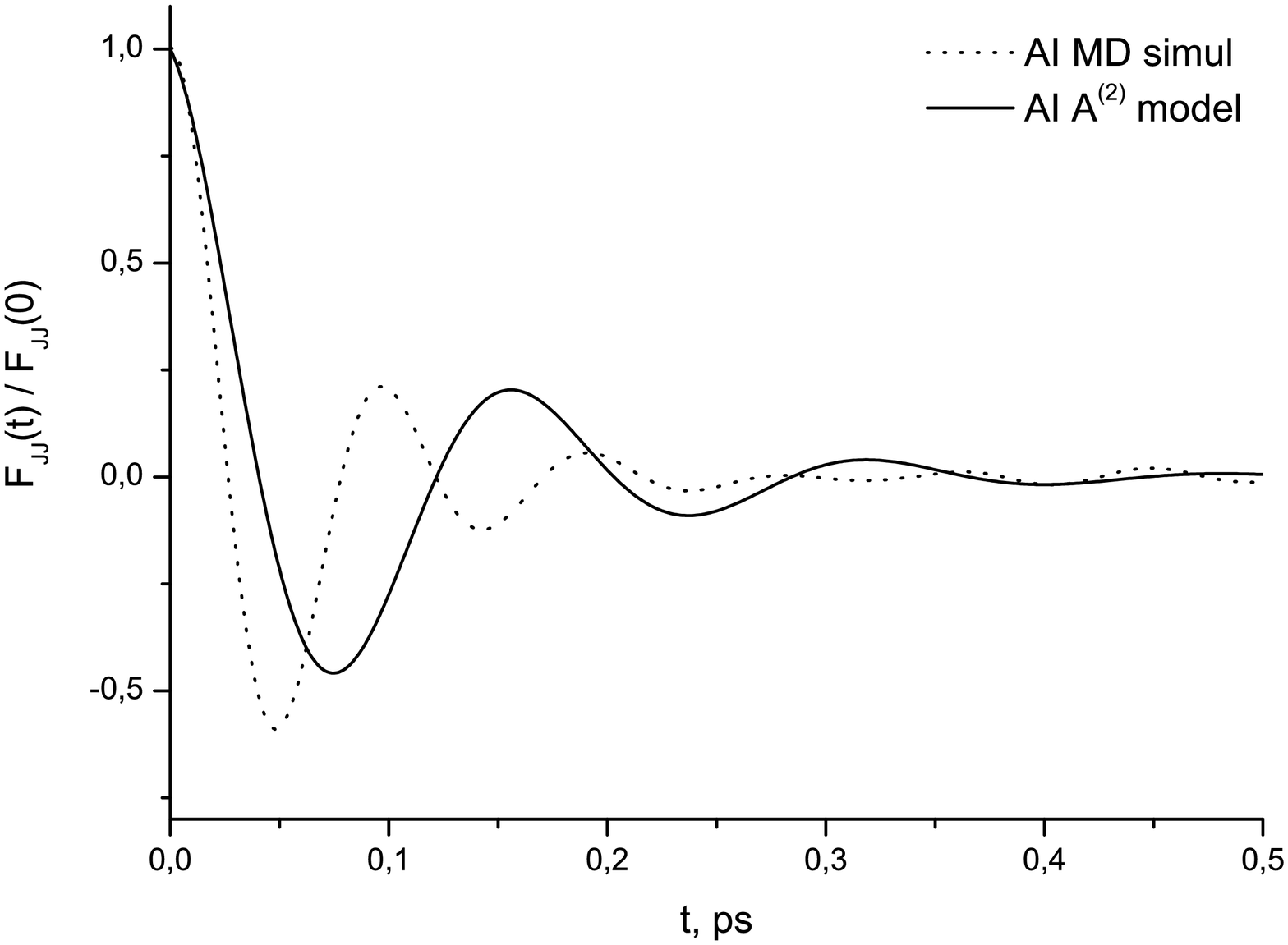}
}
\caption{Charge current normalized time correlation function for the LiBr melt at $k \sim 0.38$~{\AA} and $T=905$~K.
Left panel: Comparison of the classical MD result (dashed curve) and
the result obtained for ${\cal A}^{(2)}$ model (solid curve) within the RI model of ionic melt. Right panel: Comparison of the AI MD result (dotted curve) and
the corresponding result obtained for ${\cal A}^{(2)}$ model (solid curve). \label{fig1}}
\end{figure}

Based on the dynamic variables (\ref{1}), (\ref{4}) we can study
the influence of polarization processes on the longitudinal dynamics in ionic melts.
As it was shown earlier by means of computer simulations \cite{bryk},
the propagation of acoustic excitations is related with the fluctuations of mass current, whereas fluctuations of charge variables correspond to optic modes.
It also should be noted that  the ``$t-q$'' correlations are weak enough in ionic melts in the region of small $\vk$ \cite{bryk,43}. Taking this into account and the fact that we are interested in the influence of polarization processes on optic modes,
we will consider a simplified model
for collective dynamics which is based on the model ${\cal A}^{(2)}$ built on the dynamic variables of charge current and its time derivative only:
\begin{eqnarray}
\label{A2}
{\cal A}^{(2)}=\{ \hat{J}_{\vk}, \dot{\hat{J}}_{\vk}\}.
\end{eqnarray}
Despite the simplicity, this model qualitatively describe the considered processes.
In order to ensure that, in figure~\ref{fig1} we compare the charge current time autocorrelation function for the LiBr melt \cite{klev} calculated based on the model
(\ref{A2}) and obtained from computer simulations.
Within the ${\cal A}^{(2)}$ model the correlation function was calculated using formula
(\ref{FJJ_RI}). The details of the simulations can be found in Ref. \cite{klev}.
As we can see from the figure, the time correlation function of charge current obtained based on the set of dynamic variables ${\cal A}^{(2)}$ in general qualitatively describe the results of computer simulations by means of both classical molecular dynamics (left panel) and ab initio simulations (right panel) which takes into account polarization processes. Although quantitative agreement is not so good, we think that such a simplified model allows us to study the influence of ion polarization on optic excitations.
Comparing the left and the right panels we can also make sure that the frequency of optic modes obtained from classical molecular dynamics is more than the one obtained from AI MD.

For the NaCl, NaI, LiBr and some other melts it is typical that
the polarizability of ions of one species is much more then the other one.
That is why we consider the situation when only the negatively charged ions are polarized.
In such a case a minimal set of dynamic variables should contain the charge current, its time derivative and the variable of the induces dipole moment of the negatively charged ions:
\bea
{\cal A}^{(3)}=\{ \hat{J}^q_{\vk}\,, \  \dot{\hat{J}}^q_{\vk} \,,  \  \hat{d}_{\vk}^{-} \}.
\eea

For the sake of further simplicity of calculations, it is convenient to pass to the orthogonalized set of variables:
\bea
\label{a1}
&&\hat{a}_{1,\vk}=\frac{\hat{J}^{q}_{\vk}}{\langle \hat{J}^{q}_{\vk}\hat{J}^{q}_{-\vk} \rangle^{1/2}}\,,
\qquad
\hat{a}_{2,\vk}=\frac{\dot{\hat{J}}^{q}_{\vk}}{\langle \dot{\hat{J}}^{q}_{\vk} \dot{\hat{J}}^{q}_{-\vk} \rangle^{1/2}}\,,\\
\label{a3}
&&\hat{a}_{3,\vk}=
\frac{1}{C_3^{1/2}(k)}\left(\hat{d}_{\vk}^{-}-
\gamma(k)\dot{\hat{J}}^{q}_{\vk}
\right),
\eea
where
\bea
&& \gamma(k)=\frac{\langle \hat{d}_{\vk}^{-} \dot{\hat{J}}^{q}_{-\vk} \rangle}
{\langle \dot{\hat{J}}^{q}_{\vk} \dot{\hat{J}}^{q}_{-\vk} \rangle}, \qquad
C_3(k)=\langle {\hat{d}}_{\vk}^{-} {\hat{d}}_{-\vk}^{-} \rangle-
\gamma(k) \langle\dot{\hat{J}}^{q}_{\vk} {\hat{d}}_{-\vk}^{-}\rangle.
\eea
Calculating the flow of the dipole moment we obtain
\begin{eqnarray}
\label{flow-d}
iL_N  d^-_{\vec k}=\mathrm{w}^-_{\vec k}
+{i k}{{\Pi}}{}_{d,{\vec k}}^-\,,
\end{eqnarray}
where $\mathrm{w}^-_{\vec k}$ and ${{\Pi}}{}_{d,{\vec k}}^-$ are the longitudinal components of corresponding quantities
\begin{eqnarray*}
\vec w^-_{\vec k}=\sum_{l=1}^{N_-}\vec w_l \times \vec d_l^- e^{i\vec{k} \cdot{\vec
r}_l}, \qquad
{\mathbf{\mathbf{\Pi}}}{}_{d,{\vec k}}^-=\sum_{l=1}^{N_-}\vec p_l^- \vec
d_l^- e^{i\vec{k} \cdot{\vec r}_l}/m_{-}.
\end{eqnarray*}
Taking into account the symmetry properties and relation (\ref{flow-d})
we get the following results for $\gamma(k)$ and $C_3(k)$:
\bea
&& \gamma(k)=- \frac{i k \langle \hat{\Pi}_{d,\vk}^{-} \hat{J}^{q}_{-\vk} \rangle}
{\langle \dot{\hat{J}}^{q}_{\vk} \dot{\hat{J}}^{q}_{-\vk} \rangle}, \qquad
C_3(k)=\langle {\hat{d}}_{\vk}^{-} {\hat{d}}_{-\vk}^{-} \rangle-
\frac{k^2\langle \hat{\Pi}_{d,\vk}^{-} \hat{J}^{q}_{-\vk} \rangle^2}
{\langle \dot{\hat{J}}^{q}_{\vk} \dot{\hat{J}}^{q}_{-\vk} \rangle},
\eea
which should be considered in the long-wavelength limit.

In order to find the collective modes we should calculate the eigenvalues of the
generalized hydrodynamic matrix
$\mathbf{T}(k)=-i\mathbf{\Omega}(k)+\boldsymbol{\varphi}(k)$.
Here, the elements of the frequency matrix are expressed by the relation
$i{\Omega}_{lm}(k)=\langle\dot{\hat{a}}_{l,\vk}\hat{a}_{m,-\vk}\rangle$ ($l,m=1,2,3$).
It is easy to check that in our case the frequency matrix has the following form:
\bea
i\mathbf{\Omega}(k)=\left(
\begin{array}{ccc}
0 & \omega_0(k) & 0 \\
-\omega_0(k) & 0 & 0 \\
0 & 0 & 0
\end{array}
\right),
\eea
where
\bea
\omega_0(k)=\left(\frac{\langle\dot{\hat{J}}^{q}_{\vk}\dot{\hat{J}}^{q}_{-\vk}\rangle}
{\langle \hat{J}^{q}_{\vk} \hat{J}^{q}_{-\vk} \rangle}\right)^{1/2}\,
\eea
is expressed via the static correlation functions.
The matrix of memory function can be written as follows:
\bea
\label{24}
\mathbf{\varphi}(k)=\left(
\begin{array}{ccc}
0 & 0 & 0 \\
0 & \varphi_{22}(k) & \varphi_{23}(k) \\
0 & \varphi_{32}(k) & \varphi_{33}(k)
\end{array}
\right).
\eea
The memory functions are calculated in the Markovian approximation
$\mathbf{\varphi}_{lm}(k)=\int_0^{\infty}\langle I_{l,\vk}(t) I_{m,-\vk} \rangle dt$. They are built on the  generalized flows $I_{l,\vk}=(1 - P)iL_N \hat{a}_{l,\vk}$,
where $P \ldots = \sum_l \langle \ldots \hat{a}_{l,-\vk} \rangle \hat{a}_{l,\vk}$ is the corresponding operator projecting on the space of dynamic variables.
For a detailed study of the memory functions (\ref{24}) let us  consider the generalized flows. Taking into account Eqs.~(\ref{a1}), (\ref{a3}) we obtain
\begin{eqnarray}
\label{flows}
&&I_{1,\vk}=(1-P)iL_N{\hat{a}}_{1,\vk}=0,\\
&&I_{2,\vk}=(1-P)iL_N{\hat{a}}_{2,\vk}=\frac{1}{\langle \hat{J}^{q}_{\vk} \hat{J}^{q}_{-\vk} \rangle^{1/2}}(1-P)iL_N\dot{\hat{J}}^q_\vk=\frac{I_{J,\vk}}{\langle \hat{J}^{q}_{\vk} \hat{J}^{q}_{-\vk} \rangle^{1/2}},\\
&&I_{3,\vk}=(1-P)iL_N{\hat{a}}_{3,\vk}=\frac{1}{C_3^{1/2}(k)}
\left(I^{-}_{d,\vk}-\gamma(k)I_{J,\vk}\right).
\end{eqnarray}
Thus, we can write down:
\begin{eqnarray}
\label{kernels}
&&\varphi_{22}(k)=\varphi_{JJ}(k),\\
&&\varphi_{23}(k)=\varphi_{32}(k)=\varphi_{Jd}(k)
-\gamma(k)\frac{\langle \hat{J}^{q}_{\vk} \hat{J}^{q}_{-\vk} \rangle^{1/2}}{C_3^{1/2}(k)}\varphi_{JJ}(k),\\
&&\varphi_{33}(k)=\varphi_{dd}(k)-2\gamma(k)\frac{\langle \hat{J}^{q}_{\vk} \hat{J}^{q}_{-\vk} \rangle^{1/2}}{C_3^{1/2}(k)}\varphi_{Jd}(k)+
\gamma^2(k)\frac{\langle \hat{J}^{q}_{\vk} \hat{J}^{q}_{-\vk} \rangle}{C_3(k)}\varphi_{JJ}(k).
\end{eqnarray}
Calculating the flow of  ${\hat{J}}_{\vk}^q$
\begin{eqnarray}
\dot{\hat{J}}_{\vk}^q=J_{0,\vk} + i k  {\Pi}_{J,\vk}\,,
\end{eqnarray}
with longitudinal components of
\begin{eqnarray}
\mathbf{J}_{0,\vk} = - \sum_{a} \sum_{\bsm l,j = 1 \\l\neq j \esm}^{N_a, \, N_b}
\frac{\partial U_\mathrm{ion}}{\partial \vec{r}_l}
\left(\frac{q_a}{m_a} e^{i\vk\cdot\vec{r}_l}
- \frac{q_b}{m_b} e^{i\vk\cdot\vec{r}_j}\right),  \quad
\mathbf{\Pi}_{J,\vk} = \sum_{a} \sum_{l=1}^{N_a} \frac{q_a}{m_a^2}\vec{p}_l^a \vec{p}_l^a e^{i\vk\cdot\vec{r}_l},
\end{eqnarray}
in the long-wavelength limit we get that $\varphi_{JJ}(k)=\varphi_{q} + O(k^2)$,
where $\varphi_{q}=4\pi\sigma/\varepsilon$ ($\sigma$ is the ionic conductivity coefficient).
$\varphi_{33}(k)=\varphi_{d} + O(k)$, where $\varphi_{d}=\lim_{\vec k \to
0}\int_0^{\infty}\langle \mathrm{w}^-_{\vec k}(t)\,\mathrm{w}^-_{-\vec
k}\rangle_0\langle\hat{d}^-_{\vec k}\hat{d}^-_{-\vec k}\rangle^{-1} dt$ is the normalized
coefficient of rotational diffusion of the polarized ions. And at the end, the memory function  $\varphi_{23}(k)$ can be presented as
$\varphi_{23}(k)=\delta + O(k)$, where \linebreak $\delta=\lim_{\vec k \to
0}\int_0^{\infty}\langle J_{0,\vec k}(t)\,\mathrm{w}^-_{-\vec
k}\rangle_0[\langle\dot{\hat{J}}^{q}_{\vk}\dot{\hat{J}}^{q}_{-\vk}\rangle
\langle\hat{d}^-_{\vec k}\hat{d}^-_{-\vec k}\rangle]^{-1/2} dt$. The coefficient  $\delta$ describes the cross correlations between the processes of conductivity and dipole relaxation, which are small in the $\vk \to 0$ limit.

Thus, the generalized hydrodynamic matrix in the long-wavelength limit can be written down as follows:
\bea
\label{10}
\mathbf{T}(k)\big|_{\vk \to 0}=\left(
\begin{array}{ccc}
0 & -\omega_0 & 0 \\
\omega_0 & \varphi_{q} & \delta \\
0 & \delta & \varphi_{d}
\end{array}
\right).
\eea
For further calculations we pass to the symmetric matrix
\bea
\label{11}
\mathbf{T}=\left(
\begin{array}{ccc}
0 & -i\omega_0 & 0 \\
-i\omega_0 & \varphi_{q} & \delta \\
0 & \delta & \varphi_{d}
\end{array}
\right).
\eea
Herewith, such a transformation is linear and therefore the obtained matrix has the same eigenvalues as matrix (\ref{10}).

\subsection{Softening of optic modes due to polarization processes}

To study the effect of polarization processes on the collective excitations we
have to find eigenvalues of matrix (\ref{11}). Though it is not a problem to solve
the corresponding characteristic equation exactly, the results will be
cumbersome and unsuitable for analysis.
Instead, since the cross correlations between the charge variables and the dipole one are small, we can use  the perturbation theory for collective modes \cite{bryk3,Mryglod}.
For this purpose we divide the matrix (\ref{11}) on blocks
corresponding  to the subclasses of dynamic variables
$\{\hat{a}_{1,\vk},\hat{a}_{2,\vk}\}+\{\hat{a}_{3,\vk}\}$, where $\hat{a}_{3,\vk}$ is related with the collective variable of dipole moment (\ref{4}).

Then, we present the generalized hydrodynamics matrix
in the form
$$\mathbf{T}=\mathbf{T}_0+\delta \mathbf{T}.$$
Here, $\mathbf{T}_0$ is the generalized hydrodynamic matrix in the zero approximation
\bea\label{T_0}
\mathbf{T}_0=\left(
\begin{array}{ccc}
0 & -i\omega_0 & 0 \\
-i\omega_0 & \varphi_{q} & 0 \\
0 & 0 & \varphi_{d}
\end{array}
\right)
\eea
eigenvalues of which can be easily found, and
\bea
\label{deltaT}
\delta \mathbf{T}=\left(
\begin{array}{ccc}
~0~ & ~0~ & ~0~ \\
0 & 0 & \delta \\
0 & \delta & 0
\end{array}
\right)
\eea
is the perturbation matrix.

The eigenvalues of matrix (\ref{T_0}) correspond to the collective excitations in the zero approximation. According to Refs. \cite{bryk3,Mryglod} the first corrections equal to zero, and corrections in the second approximation are defined according to relation:
\bea
\label{deltaz}
\delta z_{\alpha}^{(2)}=\sum_{\beta}
\frac{\bar{T}_{\alpha\beta}\bar{T}_{\beta\alpha}}{z_\alpha^{(0)}-z_\beta^{(0)}},
\eea
where $\beta$ can not take values from the subclass $\alpha$, and
\bea \label{barT}
\bar{T}_{\beta\alpha}=({{\vec X}_\beta^{(0)}}^*\delta {\vec T} \,{\vec X}_\alpha^{(0)})
\eea
is the perturbation matrix in the representation of eigenvectors of matrix $\mathbf{T}_0$.

Looking for the eigenvalues of matrix (\ref{T_0}) we find collective excitations in the zero approximation%
\bea
\label{14}
z_{1,2}^{(0)}=z_{\pm}^{(0)}=\frac{1}{2}\varphi_q \pm \frac{1}{2}\sqrt{\varphi_q^2-4\omega_0^2}\, ,
\eea
which correspond to the charge variables, and
\bea
\label{15}
z_{3}^{(0)}=z_{d}^{(0)}=\varphi_d\, ,
\eea
which corresponds to the dipole relaxation.

Taking into account the fact that for the melts of interest $\varphi_q^2/\omega^2_0 \ll 1$ \cite{23,26,klev}, we obtain a simple expression for the optic modes
\begin{align}
\label{16}
&z_{\pm}^{(0)} \simeq \pm i\omega_0 + \frac{1}{2}\varphi_q = \pm i\omega_0 + \Gamma_0\,.
\end{align}
Collective modes (\ref{16}) correspond to the ${\cal A}^{(2)}$ model (\ref{A2}) of collective dynamics.
Relations (\ref{15}), (\ref{16}) describe the collective optic and the relaxation dipole excitations in the zero approximation within the simplified model ${\cal A}^{(3)}$. Taking into account the coupling between the excitations we obtain the influence of polarization processes on the optic modes.

For this purpose we calculate eigenvectors corresponding to the eigenvalues  (\ref{15}), (\ref{16}). The normalized eigenvectors can be obtained in the following form:
\begin{align}
\label{X}
&\mathbf{X}_{1,2}^{(0)}=\mathbf{X}_{\pm}^{(0)}=
\frac{1}{\sqrt{2}}\left(
\begin{array}{c}
\mp 1 \\  1 \\ 0
\end{array}
\right),
\qquad
\mathbf{X}_{3}^{(0)}=\mathbf{X}_{d}^{(0)}=
\left(
\begin{array}{c}
0 \\ 0 \\ 1
\end{array}
\right).
\end{align}

Before calculating the magnitude of correlations (\ref{deltaz}), we find that
$\bar{T}_{\pm d}=\bar{T}_{d\pm}={\delta}/{\sqrt{2}}$, and thus
\bea\label{19}
\bar{T}_{\pm d}\bar{T}_{d\pm}=\frac{\delta^2}{2}\,.
\eea

Further according to Eq.~(\ref{deltaz}) we can calculate the second correlations
to the optic excitations
\bea
\label{20}
\delta z_{\pm}^{(2)}=
\frac{\bar{T}_{\pm d}\bar{T}_{d\pm}}{z_{\pm}^{(0)}-z_d^{(0)}}=
-\frac{\delta^2}{2}\frac{\pm i \omega_0 - (\Gamma_0-\varphi_d)}
{\omega^2_0+(\Gamma_0-\varphi_d)^2}\,.
\eea
In a similar way, we  find the correction to the dipole relaxation mode
\bea
\label{21}
\delta z_{d}^{(2)}=\sum_{\alpha=\pm}
\frac{\bar{T}_{d\alpha}\bar{T}_{\alpha d}}{z_d^{(0)}-z_\alpha^{(0)}}
=\delta^2
\frac{\varphi_d-\Gamma_0}{\omega^2_0+(\Gamma_0-\varphi_d)^2}\,.
\eea
Taking into consideration the calculated corrections we can write down the expressions for collective excitations $z^{(2)}_{\alpha}=z^{(0)}_{\alpha}+\delta z^{(2)}_{\alpha}$.
Combining the results (\ref{16}) and (\ref{20}) we obtain the analytical expressions for the propagating excitations with taking into account polarization processes:
\bea\label{z-opt}
z_{\pm}^{(2)}&=& \pm i \omega_q + \Gamma_q\, ,
\eea
where
\bea
\label{im-opt}
\omega_q &=& \omega_0 \left\{1-\frac{1}{2}\frac{\delta^2}{\omega^2_0+(\Gamma_0-\varphi_d)^2}\right\}
\eea
is the frequency of optic modes and
\bea\label{re-opt}
\Gamma_q &=& \Gamma_0\left\{
1+\frac{\delta^2}{2}\frac{1+\varphi_d/\Gamma_0}{\omega^2_0+(\Gamma_0-\varphi_d)^2}
\right\}.
\eea
is the damping coefficient. From the formulas (\ref{15}) and  (\ref{21}) we find the dipole relaxation mode in the second approximation in cross correlations:
\bea\label{z-d}
z_{d}^{(2)}=D=
\varphi_d+\delta^2\frac{\varphi_d-\Gamma_0}{\omega^2_0+(\Gamma_0-\varphi_d)^2}\,,
\eea
which has the correction due to coupling with optic modes.

From the relation (\ref{im-opt}) we see that the frequency of excitations corresponding to the RI model reduces on an additive value, and therefore, taking into account polarization of the ions  within the proposed model describes the softening of the frequency of optic modes in  the long-wavelength limit. This is in a complete agrement with the results obtained by means of MD simulations within the polarization modes \cite{31} and AI MD \cite{bryk,klev}.  On the contrary the correction to the damping coefficient is positive (\ref{re-opt}) what is in consistency with the results of computer simulations as well.

\subsection{Mode contributions to the time correlation functions}

As it was mentioned above the great advantage of the GCM approach consists in the possibility to present the time correlation functions as a sum of mode contributions \cite{40} [see Eqs.~(\ref{1.11}) and (\ref{1.13})].
To calculate the corresponding amplitudes  (\ref{1.12}) we need to find the eigenvectors of matrix (\ref{11}). For this purpose let us calculate the corrections to the eigenvectors (\ref{X}) within the perturbation theory \cite{bryk3,Mryglod} according to the formula
\[
\delta \mathbf{X}_{\alpha}^{(1)}=\sum_{\beta}
\frac{\delta \bar{T}_{\beta\alpha}}{z_\alpha^{(0)}-z_\beta^{(0)}}\mathbf{X}_\beta^{(0)}.
\]
The result is as follows:
\begin{equation}
\label{deltaX}
\delta \mathbf{X}_{\pm}^{(1)}=
\frac{\delta}{\sqrt{2}}\frac{\pm i \omega_0 - (\Gamma_0-\varphi_d)}{\omega_0^2+(\Gamma_0-\varphi_d)^2} \left(\begin{array}{c}
0 \\
0 \\
1
\end{array}
\right),
\quad
%
\delta \mathbf{X}_{d}^{(1)}=
\frac{2}{\sqrt{2}}\frac{\delta}{\omega_0^2+(\Gamma_0-\varphi_d)^2}
\left(\begin{array}{c}
-i\omega_0 \\
\varphi_d-\Gamma_0 \\
0
\end{array}
\right).
\end{equation}
Combining the Eqs. (\ref{X}) and (\ref{deltaX}) we can write down the
matrix $\mathbf{X}$ composed of the eigenvectors in the first approximation in the following form:
\begin{eqnarray}
\mathbf{X}=\left(
\begin{array}{ccc}
-\frac{1}{\sqrt{2}} & \frac{1}{\sqrt{2}} & -i \omega_0 \frac{ \delta \sqrt{2}}{\omega_0^2+(\Gamma_0-\varphi_d)^2}  \\
\frac{1}{\sqrt{2}}  & \frac{1}{\sqrt{2}} & -(\Gamma_0-\varphi_d) \frac{ \delta \sqrt{2}}{\omega_0^2+(\Gamma_0-\varphi_d)^2} \\
-\frac{ \delta}{ \sqrt{2}} \frac{ i \omega_0 - (\Gamma_0-\varphi_d)}{\omega_0^2+(\Gamma_0-\varphi_d)^2}  &
\frac{ \delta}{ \sqrt{2}} \frac{ i \omega_0 + (\Gamma_0-\varphi_d)}{\omega_0^2+(\Gamma_0-\varphi_d)^2}
& 1
\end{array}
\right).
\end{eqnarray}
It is not difficult to calculate the inverse matrix $\mathbf{X}^{-1}$,
herewith, taking into account a smallness of cross correlations we use an approximated relation
\[
\frac{1}{\textrm{det}\,\mathbf{X}}  =
- \left\{ 1 - {\sqrt{2}\delta^2}\frac{\omega_0^2-(\Gamma_0-\varphi_d)^2}{\left[\omega_0^2+(\Gamma_0-\varphi_d)^2\right]^2} \right\}^{-1}
\simeq - \left\{ 1 + {\sqrt{2}\delta^2}\frac{\omega_0^2-(\Gamma_0-\varphi_d)^2}{\left[\omega_0^2+(\Gamma_0-\varphi_d)^2\right]^2} \right\}.
\]

Now according to (\ref{1.12}), we can calculate the collective mode contributions to the time correlation functions. However, in Ref. \cite{43} it was proposed to use real amplitudes instead of complex one. In such a case, an arbitrary time correlation function
(\ref{1.13}) can be presented in a slightly different form
\begin{eqnarray}
\frac{{F}_{lm}(k,t)}{{F}_{lm}(k)} = \sum_\alpha A_{lm}^{\alpha}(k)e^{-D_{\alpha}(k)t}
+\sum_{\beta}\left[ B_{lm}^{\beta}(k) \cos \omega_{\beta}(k)t + C_{lm}^{\beta}(k) \sin \omega_{\beta}(k)t \right] e^{-\Gamma_{\beta}(k)t},
\end{eqnarray}
where $A_{lm}^{\alpha}(k)$ is the amplitude from the $\alpha$-th relaxation mode, and $B_{lm}^{\beta}(k)$ and $C_{lm}^{\beta}(k)$ are the amplitudes from the $\beta$-th propagating mode with frequency $\omega_{\beta}(k)$ and damping $\Gamma_{\beta}(k)$.

In this terms we calculate the correlation function ``charge current--charge current''
\begin{eqnarray}
\label{FJJ}
\frac{{F}_{JJ}(k,t)}{{F}_{JJ}(k)} = A_{JJ}e^{-Dt}
+\left[ B_{JJ} \cos \omega_{q}t + C_{JJ} \sin \omega_{q}t \right] e^{-\Gamma_{q}t}
\end{eqnarray}
with the amplitudes
\begin{eqnarray}
A_{JJ} = \sqrt{2} \delta^2 \frac{\omega_0^2}{\left[\omega_0^2 + (\Gamma_0- \varphi_d)^2\right]^2}\,,
\end{eqnarray}
from the relaxation mode and
\begin{eqnarray}
B_{JJ} = 1 - \sqrt{2} \delta^2 \frac{\omega_0^2}{\left[\omega_0^2 + (\Gamma_0- \varphi_d)^2\right]^2}\,, \qquad
%
C_{JJ} = \sqrt{2} \delta^2 \frac{\omega_0 (\Gamma_0- \varphi_d)}{\left[\omega_0^2 + (\Gamma_0- \varphi_d)^2\right]^2}
\end{eqnarray}
from the optic modes.
As we can see, the amplitude from the effective dipole mode is a quadratic in perturbation and does not provide
a main contribution to the correlation function. The amplitudes $B_{JJ}$ and $C_{JJ}$ have the corrections due to polarization processes.

In a similar way we calculated the autocorrelation function ${F}_{\dot{J}\dot{J}}(k,t)$
\begin{eqnarray}
\frac{{F}_{\dot{J}\dot{J}}(k,t)}{{F}_{\dot{J}\dot{J}}(k)} = A_{\dot{J}\dot{J}}e^{-Dt}
+\left[ B_{\dot{J}\dot{J}} \cos \omega_{q}t + C_{\dot{J}\dot{J}} \sin \omega_{q}t \right] e^{-\Gamma_{q}t}
\end{eqnarray}
with the corresponding amplitudes
\begin{eqnarray}
A_{\dot{J}\dot{J}} = \sqrt{2} \delta^2 \frac{(\Gamma_0- \varphi_d)^2}{\left[\omega_0^2 + (\Gamma_0- \varphi_d)^2\right]^2}\,,
\end{eqnarray}
\begin{eqnarray}
B_{\dot{J}\dot{J}} = 1 - \sqrt{2} \delta^2 \frac{(\Gamma_0- \varphi_d)^2}{\left[\omega_0^2 + (\Gamma_0- \varphi_d)^2\right]^2}\,, \qquad
%
C_{\dot{J}\dot{J}} = \sqrt{2} \delta^2 \frac{\omega_0 (\Gamma_0- \varphi_d)}{\left[\omega_0^2 + (\Gamma_0- \varphi_d)^2\right]^2}\,.
\end{eqnarray}
Putting $t=0$ it is easy to verify that the zero order sum rule $A_{JJ} + B_{JJ} = 1$ is satisfied.

It should be noted that tending the magnitude of induced dipoles to zero, we immediately turn to the model of rigid ions. Herewith, the formula (\ref{FJJ}) for time correlation function $F_{JJ}(t)$ reduces to the following expression
\begin{eqnarray}
\label{FJJ_RI}
\frac{{F}_{JJ}(k,t)}{{F}_{JJ}(k)} = e^{-\Gamma_{0}t}  \cos \omega_{0}t.
\end{eqnarray}
Based on the latter the theoretical results in figure~\ref{fig1} were calculated.

\begin{figure}[!t]
\centerline{
\includegraphics[width=0.65\textwidth]{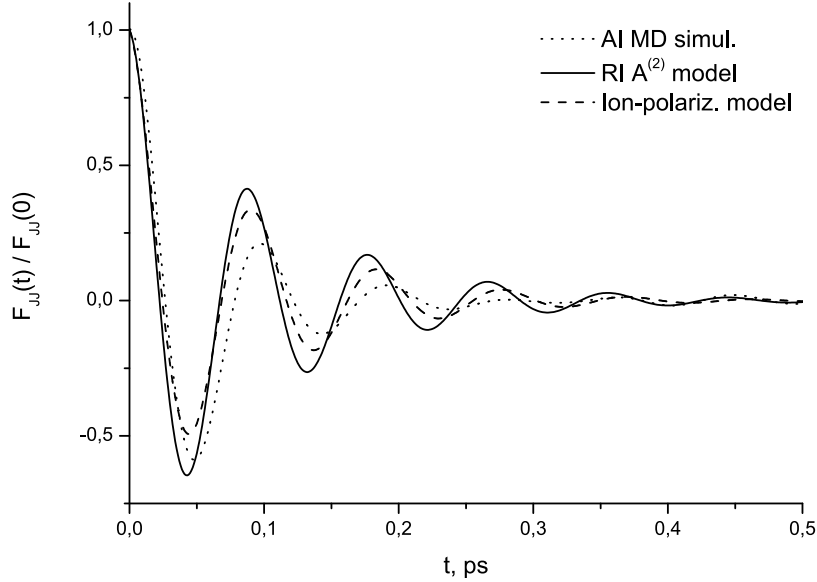}
}
\caption{Charge current normalized time correlation function for the LiBr melt at $k \sim 0.38$~{\AA} and $T=905$~K.
Comparison of the result obtained for ${\cal A}^{(2)}$ model (solid curve) within the RI model of ionic melt, the AI MD result \cite{klev} (dotted curve), result obtained based on the ${\cal A}^{(3)}$ model [see Eq. (\ref{FJJ})]  within the ion-polarization model of ionic melt (dashed curve).\label{fig2}}
\end{figure}

Figure ~\ref{fig2} shows the time correlation function ``charge current--charge current'' calculated using Eq. (\ref{FJJ}) within the ion-polarization model of ionic melt (dashed curve).
Herewith, the obtained frequency of optic modes with regard to ion polarizability $\omega_q$ is about 67~ps$^{-1}$ whereas the frequency within the RI model $\omega_0$ is more than 70~ps$^{-1}$. Thus, taking into account  polarization processes we approximate the frequency of optic modes to the value obtained from AI MD, which is about 61~ps$^{-1}$ (see Ref. \cite{klev}).
We also observe a small increase of the damping coefficient from 10~ps$^{-1}$ in RI model to 11~ps$^{-1}$ within the considered model.
As we can see from the figure, the obtained result displays the softening of the frequency of optic modes.
Taking into account polarization processes we make the theoretically calculated curve closer to the result of AI MD. The simplified model which we considered catches main characteristics only, however it allows us to obtain analytical result as well.
It worth noting, that accuracy of calculations can be improved by extending the set of dynamic variables, in particular considering the charge density fluctuations.

\section{Summary}

Within the ion-polarization model of ionic melt we considered a simplified model of collective dynamics which allowed us to describe the optic excitations in the system with taking into account the polarization processes. We studied the case when only the negatively charged ions are supposed to be polarized since the polarizability of cations is much less for many melts. In the paper the analytical expressions for the propagating optic modes and the relaxation dipole one are obtained. By means of the perturbation theory the coupling between optic and dipole modes is taken into account.
Despite the model is simple enough it allows to describe the softening of the frequency of optic modes due to the effects of ion polarization in the long-wavelength limit in comparison to the rigid-ion model. On the contrary the damping coefficient of optic modes increases. We also calculated the contributions caused by the polarization processes to the time correlation functions, in particular to the function ``charge current--charge current'' related with the conductivity of the system.
We also carried out a numerical calculation of the correlation function mentioned above, in particular we showed that the theoretically obtained charge current autocorrelation function approximates to the result of AI MD in comparison with the results of the ${\cal A}^{(2)}$ rigid-ion model and classical molecular dynamics.
Therefore, the proposed model of collective dynamics allows us to correctly take into account polarization processes.

The proposed model can be generalized. In particular, in our calculations we were interested in the  region of $\vk \to 0$, however,  taking into account
the higher terms in $\vk$ in the generalized hydrodynamic matrix permits to expand the results beyond the long-wavelength limit. The considered model can also serve as a basis for further extensions, in particular, to study the case with both anions and cations to be polarized.
One more extension consists in consideration the charge density fluctuations. Such a more complete model is very interesting because
it permits studying the charge (concentration) dynamic structure factor with taking into account the polarization processes, what is of particular interest \cite{31,klev}.

\section*{Acknowledgments}

The authors want to thank Taras Bryk for providing the data of his computer simulations.


\end{document}